\documentclass[aps,prl,twocolumn,showpacs,superscriptaddress,groupedaddress]{revtex4}  
\usepackage{graphicx}  
\usepackage{dcolumn}   
\usepackage{bm}        
\usepackage{amssymb}   

\begin{document}

\preprint{APS/MA-FR-JB}

\title{THE "PLANCKONIONS"}

\author{Mofazzal Azam}
\email{mofazzal.azam@gmail.com}
\affiliation{Centre for Theoretical Physics, Jamia Millia Islamia, New Delhi-110025, India}

\author{Farook Rahaman}
 \email{rahaman@associates.iucaa.in}
\affiliation{Department of Mathematics, Jadavpur University, Kolkata-700032, India}

\author{M Sami }
 \email{ samijamia@gmail.com}
\affiliation{Centre for Theoretical Physics, Jamia Millia Islamia, New Delhi-110025, India}

\author{ Jitesh R Bhatt }
 \email{jiteshbhatt.prl@gmail.com}
\affiliation{Physical Research Laboratory, Theory Division, Ahmedabad 380 009, India}

\date{\today}

\begin{abstract}

We consider a spherically symmetric stellar configuration
 with a density profile $\rho(r)=\frac{c^2}{8\pi G r^2} $. This configuration satisfies the Schwarzchild black hole condition
$\frac {2GM}{c^2 R}=~1$ for every $ r =R $. We refer it as "Planckonion".  The interesting thing about the Plankonion
is that it has an onion like structure. The central sphere with radius of the Plank-lenght
$ L_p=\sqrt{(\frac {2\hbar G}{c^3})}$ has one unit of the Planck-mass $M_p=\sqrt {(\frac {c\hbar}{2G})}$.
 Subsequent spherical shells of radial width $L_p$ contain
exactly one unit of $M_p$. We study this stellar configuration using  Tolman-Oppenheimer-Volkoff equation
and show that the equation is satisfied if pressure $P(r)=-\rho(r)$.
On the geometrical side, the space component of the metric blows up at every point. The time component of the metric is zero inside the star but only in the sense of a distribution (generalized  function).
The Planckonions
 mimic some features of  black holes  but avoid
 appearance of central singularity
 because of the violation of  null energy conditions.

\end{abstract}


\maketitle

\setlength{\abovedisplayskip}{3pt}
\setlength{\belowdisplayskip}{3pt}

\newpage
 Black holes have been discussed extensively in scientific monographs\cite{wald,pois,ell}  as well as popular science literature captivating the imagination of public at large \cite{ruf,haw}. Most of the time these discussions touch upon the geometric aspect of the black hole - its matter/energy content gets very little attention. What is emphasized is that  there exists  a central singularity and the event horizon and, because of "no hair"  theorem, the rest can be left as such. On the other hand, the central premise of the General  Relativity (GR) tells
us that the space-time geometry "in" and "around"  a material object is fixed by the distribution of  matter itself.
This suggests that we must understand a bit more about the state of matter/energy content to fully appreciate the black holes.
\par In the early days the interest in black hole within the theoretical physics community was more of academic in nature- it was concerned primarily with the issue of logical completion of GR . It remained as such for a very long time. However, the scenario has undergone drastic change after the direct detection of gravitational radiation on 14th September,  2015 from the merger of a black hole with its binary black hole companion \cite{ligo} .
Black hole binary of masses $(36.0\pm4.0) ~M_{\odot} $ and $(29.0\pm4.0)~M_{\odot} $ merged into a single black hole of mass $(62.0\pm4.0)~M_{\odot} $ and emitted gravitational radiation of energy $(3.0\pm 0.5)~M_{\odot} $. Gravitational radiation is quadrupolar in origin, and thus, in the binary merger process there is a transient state with quadrupolar moment which is emitted as gravitational radiation, and finally, the merged stellar configuration settles down into a single spherically symmetric configuration. In this process there is reorganization of the space-time geometry, and it is expected that there is corresponding reorganization of matter/ energy content of the stellar configuration. Thus it is important to understand the state in which matter is organized in black
hole as well as its precursor. The equation of state in thermodynamics provides the glimpse of the microscopic state of matter at the
 macrospopic scale. We attempt to find such an equation of state.

\par In General Relativity  the relationship among thermodynamic variables such as pressure,  density and mass for spherically symmetric stellar configuration is given by Tolman-Oppenheimer-Volkoff ( TOV) equation \cite{oppen,tolman,wein}. In this letter, we investigate the TOV equation for a density profile which mimics a black hole . We have called such a stellar configuration a "Planckonion"  for reason explained in the text below. A spherically symmetric stellar configuration with a density profile,
\begin{equation}
\rho(r)=\frac{ c^2}{ 8\pi G r^2},
\end{equation}

satisfies the Schwarzschild black hole condition \[ \frac {2GM}{c^2 R}=1,\] for every $r=R$. The interesting thing about this stellar configuration is that it has an onion like structure. The central sphere of radius $r =L_p$, where, $ L_p=\sqrt {\frac {2\hbar G}{c^3}}$ is the the Planck Length contains exactly one unit of Planck Mass, $M_p=\sqrt {\frac {c\hbar}{2G}}$. Subsequently, each spherical shell of radial width equal to Planck length, $ L_p $, contains exactly one unit of Planck mass, $M_p $. This is the stellar configuration we call a "Planckonion".

The mass $ M(R) $  contained within a sphere of radius $ R $ with the density profile,
\begin{equation}
\rho (r)=\frac {c^2}{8\pi Gr^2},
\end{equation}
is given by
\begin{equation}
M (R)=\int_{0}^{R} 4\pi r^2 \rho (r) dr=\frac {c^2 R}{2G}.
\end{equation}
Thus, the Schwarzschild black hole condition $ \frac{ 2GM}{c^2 R}=1$ is satisfied for any $ r=R $.
Mass contained in the central spherical region of radius $ L_p $,

\begin{equation}
M (L_p)=\frac{ c^2  L_p}{2G}=\frac{ c^2}{2G}\times \sqrt {\frac{ 2\hbar G}{c^3}}=\sqrt {\frac{c\hbar}{2G}}=M_p.
\end{equation}

Mass contained in a spherical shell of radial width $ L_p $ between spheres of radius $ r $ and $ r+L_p $,
\begin{eqnarray}
M (shell)=M (r+L_p)-M (r) \nonumber\\ =\frac{c^2(r+L_p)}{2G}-\frac{c^2 r}{2G}=\frac{c^2 L_p}{2G}=M_p.
\end{eqnarray}

This clearly shows that the stellar configuration has an onion like structure with a Planck mass at the Planck length size centre, and subsequently, a Planck mass in each spherical shell of radial width equal to the Planck length. Note that the Planckonion can not be compressed further because it satisfies the Schwarzschild black hole condition for each $ r $. The maximum mass that can be put within a sphere of radius $ R $, is $M=\frac{c^2}{2G} R$. If we increase the  mass content of sphere, its radius will also increase. This is implied by the Schwarzschild condition. Thus the Planckonions can not have pulsation.\\
\par We study this configuration using the TOV equation. To avoid the small denominator problem, we consider a slightly altered density profile,
\begin{eqnarray}
\rho (r)=\frac {\eta c^2}{8\pi G r^2},
\end{eqnarray}
and then consider the $\eta \rightarrow 1$.  From onwards, we take $ c =1$.
\par Before we write the TOV equation, we introduce the following notations,
\begin{eqnarray}
U (r)=- r^2 P'(r) \nonumber\\
V (r)=1-\frac {2G M (r)}{r} \nonumber\\
X (r)=GM (r)\rho (r) \nonumber\\
Y (r)=1+\frac{P (r)}{\rho (r) }\nonumber\\
Z (r)=1+\frac {4\pi r^3 P (r)}{M (r) }.
\end{eqnarray}

In terms of these variables the  TOV equation is,

\begin{eqnarray}
U (r) V (r)=X (r) Y (r) Z (r).
\end{eqnarray}

To proceed further, we assume an equation of state of the form $ P (r)=\omega \rho (r) $, where  the equation state parameter, $\omega $ is a constant independent of $ r $. We will find $\omega $ by solving the TOV equation. It is easy to see that for $\rho(r)=\frac {\eta}{8\pi Gr^2} $,  the mass contained within a sphere of radius $ r $  is, $M (r)=\frac {\eta r}{2G}$. Thus, by substituting  these values and  using the equation of state we have
\begin{eqnarray}
V (r)=1-\eta, ~~~Y (r)=Z (r)=1+\omega.
\end{eqnarray}

Substituting these values  in the TOV equation and  after doing some algebraic manipulations, we find
\begin{eqnarray}
4\omega \left(~~\frac {1}{\eta}-1\right)=(1+\omega)^2.
\end{eqnarray}

Introducing a new parameter $\frac{\epsilon}{2}=\frac {1}{\eta}-1$, we find
\begin{equation}
2\omega \epsilon=(1+\omega)^2.
\end{equation}

Thus the limit, $\eta\rightarrow 1$ corresponds to $\epsilon\rightarrow 0$.
The equation state parameter $\omega$ can be obtained  from the quadratic equation.
\begin{equation}
\omega^{\pm}=-1+\epsilon\pm \sqrt {\epsilon^2-2\epsilon}.
\end{equation}

For $\epsilon=8/3$ ($\eta=3/7$), we find, \[ \omega^{-}=1/3~~ and ~~\rho (r)=\frac {3}{56\pi G r^2} .\] In this case the equation state is given by $ P=\frac{1}{3}\rho$.
 This is the equation of state which is used for relativistic neutrons in a neutron star \cite{oppen,wein}. \\
To understand the geometric aspect, let us recall that in GR for a spherically symmetric stellar configuration the proper time is given by,
\begin{equation}
d\tau^2= B (r) dt^2-A (r) dr^2-r^2 (d\theta^2+\sin^2 \theta d\phi^2).
\end{equation}
The pressure and density appearing in TOV equation are related to time and space components of the metric
 by the equations \cite{oppen,wein} ,
\begin{equation}
A (r)=\frac {1}{1-\frac {2GM}{r}},
\end{equation}
and
\begin{equation}
\frac {-2P'(r)}{P (r) +\rho (r)}=\frac {B'(r)}{B (r)}.
\end{equation}

For neutron stars it is easy to find that \[  A (r)=4/7~~ and ~~ B (r)\approx 1/\sqrt {\rho (r)} .\] This shows that time, as expected, considerably slows down
in the central region where the density is high. However, near the boundary the matter density is lower and we must use the non-relativistic
equation of state.
The inverse square density profile, $\rho(r)= \frac{3}{56\pi G r^2} $, with equation of state, $ P=\frac{1}{3} \rho $, solves the TOV equation but fails to describe the neutron stars accurately.\\
\par When $\epsilon < 2$, the equation state parameter $\omega $ develops an imaginary part. Equation of state
with such an $ \omega $  for any stellar
configuration should correspond to non-static nature of its thermodynamic state. However, we are interested in the $\epsilon \rightarrow 0$ and would like
to investigate whether we can arrive at some sensible limit without bothering about the thermodynamics. Thus "non-static" for us will simply refer to contraction . From the TOV equation, it is easy to find that in this limit density, pressure and its derivative do not have  imaginary part, and equation of state
is simply $ P(r)= -\,\rho(r) $. Before, we comment on this equation of state,  we would like to discuss what happens to the components of metric tensor in
this limit. It is clear that the space component of the metric, $ A (r)\rightarrow \infty $. However, the calculation of the time component of the
metric, $ B (r) $, is a bit more complicated. From Eq. (14), using the equation of state $ P (r)=\omega \rho (r) $, and equation Eq. (10) above,
we obtain,
\begin{equation}
( \ln B (r))'=-\frac {2 P' (r)}{P (r)+\rho (r)}=-\frac {1+\omega}{\epsilon}(\ln\rho (r))'.
\end{equation}
Now using using Eq. (11), for $\epsilon  $  very very small,
we get,
\begin{equation}
(\ln B (r))'= -\left(1\pm  i \sqrt {\frac {2}{\epsilon}} \right)~~ (\ln \rho (r))'.
\end{equation}
Considering only the positive sign within the bracket, we can write down the solution as,
\begin{equation}
\frac {B (r)}{B (R)}=\frac {\rho (R)}{\rho (r)}\left ( \frac {\rho (R)}{\rho (r)}\right )^{ i\sqrt {\frac{2}{\epsilon}}}.
\end{equation}
Here $ r $ is any interior point of the star and $ R<\bar{R} $ is arbitrary close to $\bar{R} $, the point at boundary. For the density profile in this letter,
\[ \rho (R)/\rho (r)=r^2/R^2.\]
 Thus $ B (r) $ is complex, and is a highly oscillatory  function of
$ r $   in the limit of $\epsilon\rightarrow 0$.
 In fact, it is non-analytic in $\epsilon $ (at $ \epsilon=0$).
Clearly, the strong limit for $\epsilon\rightarrow 0 $ does not exist. But the weak limit (in the sense of distribution/generalized function) exists, and is equal
to zero in the interior of the star (for $ r <R $). This can easily be seen as follows. First of all, note that \cite{note}
\begin{eqnarray}
lim_{\alpha\rightarrow \infty}  e^{ i \alpha x} : =\delta (x)\nonumber
\end{eqnarray}
Thus Eq. (18) in the limit of $\epsilon\rightarrow 0$ becomes,
\begin{eqnarray}
\frac {B (r)}{B (R)} : =\frac{\rho (R)}{\rho (r)}\delta \Big( ln\frac {\rho (R)}{\rho (r)}\Big)
\end{eqnarray}
It follows that in the interior of the star ($ r <R $), time component of the metric $ B (r) $ can be taken to be zero in the sense of a distribution
(generalized function).
However, this component of the metric at boundary does not seem to be well defined.
\par Regarding the negative pressure, a few comments are in order.  It has been suggested in some publications that negative pressure can possibly be understood as tension \cite{kar}. In our context we understand
it as follows. The gravitational energy of a sphere of mass "$ M $" and radius "$ R $"  is of the order of, "$-\frac{GM^2}{R} $". For black hole of radius
 $ R_{BH}=\frac{2GM}{c^2} $, this will be of the order of,  "$-Mc^2~$". Thus if this energy were radiated away before the black hole formation, the black hole would be left with a small fraction of the  inertia " $ M $ ".
 It would be tempting to assume that this energy is utilised in exciting dynamical
graviton modes, the other matter particle degrees of freedom being non-dynamical at this stage.
As in the case of protons where large part of its mass is concentrated in gluon field, we expect that the large part of  black hole mass
 is concentrated in graviton field. As per Dvali-Gomez conjecture \cite{dvali} this energy is dominated by gravitons of system size wavelength. However,
there will also be gravitons of shorter wavelength but of lesser weight in the spectrum. The gravitons of system size wavelength in a black hole are at their maximum packing,  and  are enormously
large  in number. They are also very weakly interacting. As per Dvali-Gomez picture, these system size gravitons within the black hole undergo Bose condensation. Gravitons of shorter wave  lengrh (higher momentum) would create local strain in the condensate.
 Bose condensate is macroscopic in size  but it is a single quantum
state and will certainly exhibit tension under local stretching.
 At macroscopic level force in gravity is  long range, central and attractive, and
falls off as inverse square, and thus, we expect a differential
stretching in the condensate. We speculate that the emergent tension and thus negative pressure is due to such a mechanism. This picture is equally applicable to Planckonions. \\

\par Our aim in this letter  has been to try to understand the black holes without the appearance of central singularity. We found that, in some sense,
the "Planckonions" with a density profile, $\rho (r)=\frac {c^2}{8\pi G r^2} $ serves the purpose. It leads to a equation of the form
$ P (r)=- \rho (r) $, and thus, violates the null energy condition and prevents the appearance of central singularity \cite{pen,rub} .
The space components of the
metric is infinite at every point including the boundary. The time component  is zero within the star but only in the restricted sense of weak convergence
as explained in the text. However at the boundary, this component of the metric remains uncertain. From the analysis in this letter, it is clear that we can not yet designate the black holes
as "Planckonions"  but some of the features found in this work look promising.



\newpage

\begin {thebibliography}{99}

\bibitem{wald}  Robert M. Wald, General Relativity, Chicago University Press, Chicago, 1984

\bibitem{pois}  Eric Poisson, A relativist's toolkit:The mathematics of black hole mechanice, Cambridge University Press, Cambridge,  2004

\bibitem{ell}    S.W. Hawking and G.F.R Ellis, The large scale structure of space-time, Cambridge University Press,  Cambridge,  1973

\bibitem{ruf}   R. Ruffini and J.A. Wheeler, Physics Today {\bf 24} (1971) 30-41

\bibitem{haw}   S.W. Hawking,  A brief history of time, Bantam Dell Publishing Group, New York,1988

\bibitem{ligo}    B.P. Abbot et al (LIGO Scientific Collaboration and Virgo Collaboration) Phys. Rev. Lett  {\bf 116} (2016) 061102

\bibitem{oppen}   J.R. Oppenheimer and G.M. Volkoff,  Phys. Rev. {\bf 55} (1939) 374

\bibitem{tolman}   R.C. Tolman,   "Relativity, Thermodynamics and Cosmology" ,  Oxford 1934

\bibitem{wein}   S. Weinberg,  Gravitation and Cosmology, John Wiley and Sons, Inc 1972

\bibitem{note}   This equation means that the highly oscillatory function induces a singular measure concentrated at $x=0$. On the face of it, this representation of the delta function is slightly nonstandard. \\
   I M. Gelfand and G.E. Shilov,  Generalized Functions Vol 1 , Academic Press, New York 1964


\bibitem{pen}    R. Penrose, Phys. Rev. Lett. {\bf 14} (1965) 57

\bibitem{rub}    V.A. Rubakov, The null energy condition and its violation,  Physics-Uspekhi {\bf 57 (2)} (2014) 128-142;  arxiv: 1401.4024

\bibitem{kar}  Karen Wright, The Physics of negative pressure,  Discover Magazine, March 2003
\bibitem{dvali}  G. Dvali and C. Gomez, "Black Hole's Quantum N-Portrait" ,  Fortsch.  Phys. 61 (2011) 742-767;  arXiv:1112.3359 (hep-th)

\end {thebibliography}
\end{document}